\documentclass[a4paper]{jpconf}
\usepackage{graphicx}
\usepackage{amsmath}
\usepackage{amsfonts}
\begin{document}
\title{Supersymmetric infinite wells and coherent states}

\author{M-A Fiset$^1$ and  V Hussin$^{2,3}$}

\address{$^1$Department of Physics,
McGill University, Montr\'eal, QC H3A 2T8, Canada.}
\address{$^{2}$D\'epartement de math\'ematiques et de statistique, Universit\'e de Montr\'eal, C. P. 6128, Succ. Centre-ville, Montr\'eal (Qu\'ebec) H3C 3J7, Canada.}
\address{$^3$ Centre de recherches math\'ematiques, Universit\'e de Montr\'eal, C. P. 6128, Succ. Centre-ville, Montr\'eal (Qu\'ebec) H3C 3J7, Canada.}

\ead{marc-antoine.fiset@mail.mcgill.ca, veronique.hussin@umontreal.ca}

\begin{abstract}
Gaussian Klauder coherent states are discussed in the context of the infinite well quantum model, otherwise known as the particle in a box. A supersymmetric partner system is also presented, as well as a construction of coherent states in this new system. We show that these states can be chosen, in both systems to have many properties usually expected for coherent states. In particular, they yield highly localised wave packets for a short period of time, which evolve in a quasi-classical manner and which saturate approximately Heisenberg uncertainty relation. These studies are elaborated in one- and two-dimensional contexts. Finally, some relations are established between the gaussian states being mostly used here and the generalised coherent states, which are more standardly found in the literature.
\end{abstract}

\section{Introduction}
The study of coherent states in quantum mechanics is a subject very well documented \cite{Schro, Glauber, KS, Perelomov, CT, Gazeau}. These states have been approached from different points of view and we can cite, for example, their definition as eigenstates of an annihilation operator and as minimium uncertainty states. They have also been defined for a large amount of solvable quantum systems and their properties have been analysed \cite{Gazeau, Dong2, Angelova}.

One approach, that has not deserved much attention until now, is the one using the so-called gaussian Klauder coherent states \cite{Fox-Choi2000}. They are given as a superposition of energy eigenstates that leads to a good localisation in the phase space of the system under consideration. For the harmonic oscillator, these last states are a good approximation of the usual coherent states. It has been shown that it is also the case for the infinite well and the Morse systems \cite{Angelova, Fox-Choi2000,Fox-Choi2001, FH}. It means that they are good states for investigating physical properties of the systems under consideration (behaviour of the position and momentum observables and uncertainty relation, for example).

Even if most of the works were focussing on one dimensional (1D) systems, some results have been given in the two dimensional (2D) case \cite{Fox-Choi2001}. In particular, the 2D infinite well, having a quadratic degenerate energy spectrum, has attracted some attention and the construction of coherent states had to be adjusted \cite{Fox-Choi2001, DH} .

All these considerations have been extended using a supersymmetric (SUSY) approach \cite{FHR}. Indeed, SUSY partner Hamiltonians have been constructed in 1D and 2D using intertwining relations and factorisation methods \cite{FHR, FS, Ioffe}. These new Hamiltonians are shown to be deeply related to the original ones. The coherent states systems are as well closely related to the original ones.

In this work, we consider the infinite well quantum system both in 1D and 2D and their SUSY partners. We then construct the Gaussian Klauder coherent states (GCS) and investigate their properties. In Section 2, we show how the SUSY partners of the 1D infinite well are obtained. New potentials are thus produced that depend on two parameters (one is an integer and the other one is a real number). The complete energy spectra of the new Hamiltonians are obtained. The GCS are then constructed and their properties are exhibited. In Section 3, we generalise the preceding approach to 2D where the SUSY partners exhibit more freedom. Complete energy spectra are also produced. We generalise the definition of GCS to this context. We discuss as well another type of coherent states, the so-called generalised coherent states (GeCS). Because of the degeneracy of the energy spectrum, the usual definition of those states as the eigenstates of an annihilation operator has to be adjusted. We finish the paper with some conclusions and questions for future work.

\section{The 1D infinite well and SUSY partners }

\subsection{Description of the models}

Let us first set our notational convention concerning the infinite square well  \cite{Fox-Choi2000} to be used throughout this work. A particle of mass $M$ is subject to a potential taken to be
\begin{equation}
V(x)=\begin{cases}0,&0<x<\pi\\\infty,&\text{otherwise}.\end{cases}
\end{equation}
The stationary eigenstates and the discrete energies of this system are
\begin{equation}
\psi_n(x)=\sqrt{\frac{2}{\pi}}\sin{nx},\quad\quad E_n=\frac{\hbar^2}{2M}n^2, \quad n=1,2,\dots
\label{eigenIW}
\end{equation}
In the following, we will use dimensionless units, setting $\hbar=1$, $M=1/2$, such that the Hamiltonian is $H_x=- \frac{d^2}{dx^2}+V(x)$.

SUSY partners of the infinite well have been constructed  \cite{FHR} starting from usual intertwining relations involving the supercharges $Q_{x}$, $Q_{x}^\dagger$. These can be defined, in particular, as differential operators of second order. Let us summarize the results.

Starting from the Hamiltonian $H_x$, a SUSY partner Hamiltonian $\tilde{H}_x=- \frac{d^2}{dx^2}+\tilde{V}(x)$ is obtained from the relations
\begin{equation}
\tilde{H}_x Q_{x}=Q_{x} H_{x},\quad Q_{x}^\dagger \tilde{H}_{x} = H_{x} Q_{x}^\dagger.
\label{intertwining}
\end{equation}
For the infinite well, with $V(x)=0$ in the domain $x\in ]0,\pi[$, internal consistency constrains the supercharges to take the form
\begin{equation}
Q_x=\frac{d^2}{dx^2}+ \eta(x) \frac{d}{dx}+\epsilon +\frac12 ( \eta^2(x) - \eta'(x) )
\end{equation}
and 
\begin{equation}
Q^\dagger_x=\frac{d^2}{dx^2}- \eta(x) \frac{d}{dx}+\epsilon +\frac12 ( \eta^2(x) -3 \eta'(x) ),
\end{equation}
where $\epsilon$ is an arbitrary constant. The function $\eta(x)$ satisfies (in the so-called confluent case \cite{FS}):
\begin{equation}
2 \eta(x) \eta''(x)-(\eta'(x))^2-4 \eta^2(x)\eta'(x)+\eta^4(x)+4\epsilon \eta^2(x)=0
\label{etasol}
\end{equation}
and the new potential is given as
\begin{equation}
\tilde{V}(x)=2 \eta'(x).
\end{equation}
Moreover, the products $Q_{x}^\dagger Q_{x}$ and $Q_{x}Q_{x}^\dagger$ are respectively polynomials in $H_{x}$ and $\tilde{H}_{x}$:
\begin{equation}
Q_{x}^\dagger Q_{x}=(H_{x}-\epsilon)^2, \quad Q_{x}Q_{x}^\dagger=(\tilde{H}_{x}-\epsilon)^2.
\label{factorizations}
\end{equation}
The resolution of (\ref{etasol}) leads to admissible solutions for $\epsilon=k^2$ with  $k=1,2,...$ A particular energy $\epsilon=E_k$ within the original spectrum thus needs to be chosen for the supersymmetry. We get \cite{FS}
\begin{equation}
\eta(x;k,\omega)=\frac{4k\sin^2(kx)}{\sin(2kx)+2k(\pi\omega-x)},
\end{equation}
where $\omega$ is an arbitrary constant.
The corresponding potentials are given as
\begin{equation}
\tilde{V}(x;k,\omega)=\begin{cases} \frac{32k^2\sin(kx)[\sin(kx)+k(\pi\omega-x)\cos(kx)]}{[\sin(2kx)+2k(\pi\omega-x)]^2},&0<x<\pi\\\infty,&\text{otherwise}.\end{cases}
\label{potential}
\end{equation}
These potentials can be shown to be non singular if  $\omega\in ]-\infty,0[\ \cup\ ]1, \infty[$. Two instances are illustrated on figure 1, one for each of the disconnected subsets of the parameter space of $\omega$, i.e.  $\omega=2$ and $\omega=-1$. We thus observe that changing the sign of $(\omega-1/2)$ reverses this potential about its vertical axis. Decreasing the magnitude of $|\omega-1/2|$ accentuates the departure from a sinusoidal potential. The parameter $k$ controls the amount of oscillations.

\begin{center}
\includegraphics[scale=0.52]{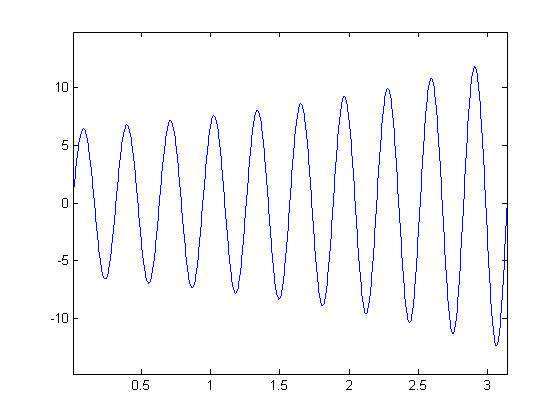}
\includegraphics[scale=0.52]{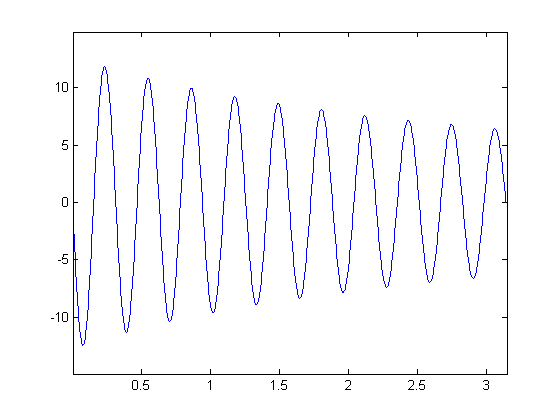}

\small{Figure 1 - SUSY potential $\tilde{V}(x;k=10,\omega)$ as a function of $x$, for (left) $\omega=2$ and (right) $\omega=-1$.}
\end{center}

Notice the striking symmetry under the $\mathbb{Z}_2$ action $\omega\rightarrow 1-\omega$, $x\rightarrow \pi-x$. This symmetry is accidental in the sense that it is not a consequence of the SUSY algebra, nor the imposed intertwining relations \eqref{intertwining}. Still, it will have interesting effects on the coherent states built out of SUSY potential functions, as we shall discuss in section 2.3.

The normalised SUSY eigenstates $\tilde{\psi}_n(x)$ are obtained from the intertwining relations (\ref{intertwining}) and the expressions (\ref{factorizations}):
\begin{equation}
\tilde{\psi}_n(x;k,\omega)= (\epsilon-E_n)^{-1}Q_x\psi_n(x),\quad     (n\neq k).
\label{eigenSUSYIW}
\end{equation}
More explicitly, we get
\begin{equation}
\tilde{\psi}_n(x;k,\omega)=\sqrt{\frac{2}{\pi}}\frac{\sin(nx)[\sin(2kx)(n^2+k^2)+2k(\pi\omega-x)(n^2-k^2)]-4nk\cos(nx)\sin^2(kx)}{(n^2-k^2)[\sin(2kx)+2k(\pi\omega-x)]}\label{psi_n}
\end{equation}

For $n\neq k$, they are physical states, i.e. they are normalisable and such that $\tilde{\psi}_n(0;k, \omega)=\tilde{\psi}_n(\pi;k,\omega)=0$, since $\eta(0;k,\omega)=\eta(\pi;k,\omega)=0$. The corresponding energies are $E_n=n^2$ as in the original case. 

For $n=k$, since $\epsilon=E_k=k^2$, this procedure does not yield $\tilde{\psi}_k(x)$ because $Q_x{\psi}_k(x)=0$. The completeness of the spectrum of $\tilde{H}_x$ has been investigated \cite{FS} and a single additional state was found. For the sake of completeness, we now explain how this missing state $\tilde{\psi}_k(x)$ could be obtained. It is, in fact, the simultaneous solution to the system of equations
\begin{equation}
Q_x^{\dagger}\tilde{\psi}_k(x)=0, \quad H_x \tilde{\psi}_k(x)=\epsilon\tilde{\psi}_k(x),
\end{equation}
that reduces to a first order differential equation on $\tilde{\psi}_k(x)$:
\begin{equation}
- \eta(x;k,\omega) \frac{d\tilde{\psi}_k(x)}{dx}+(\tilde{V}(x;k,\omega) +\frac12 ( \eta^2(x;k,\omega ) -3 \frac{d\eta(x;k,\omega)}{dx} ))\tilde{\psi}_k(x)=0.
\label{missingequ}
\end{equation}

From \cite{FS}, we know that $\eta'(x)=\eta^2(x)+2\beta(x) \eta(x)$, with $\beta(x)$ solving the Riccati equation $\beta'(x)+\beta^2(x)=-\epsilon$. Equation (\ref{missingequ}) thus becomes
\begin{equation}
\frac{d\tilde{\psi}_k(x)}{dx}=(\eta(x;k,\omega)+\beta(x;k))\tilde{\psi}_k(x)
\label{missingequ1}
\end{equation}
and the normalised solution of energy $\epsilon=E_k=k^2$ is given as 
\begin{equation}
\tilde{\psi}_k(x;k,\omega)=\sqrt{\frac{2}{\pi}}\sin(kx)\frac{2\pi k\sqrt{\omega(\omega-1)}}{\sin(2kx)+2k(\pi\omega-x)}\label{psi_k}.
\end{equation}
It is normalisable and such that $\tilde{\psi}_k(0,k,\omega)=\tilde{\psi}_k(\pi,k,\omega)=0$. With this additional state the spectrum of $\tilde{H}_x$ is thus complete.

\subsection{Gaussian Klauder coherent states for the infinite well}

As mentioned in the introduction, the gaussian Klauder coherent states (GCS) can be built for many different systems as a special superposition of energy eigenstates in order to get a reasonably well localised probability density distribution for a short period of time \cite{Fox-Choi2000}. They have proven to be relevant for the study of the harmonic oscillator and the infinite well. For this last system, we recently \cite{FH} formalised the relation between those states and the generalised coherent states (GeCS), constructed as eigenstates of an annihilation operator of the system under consideration. We have shown in particular that the GCS can be chosen to reproduce approximately the GeCS in some specific area of their parameter space. In this section, we thus deal only with those GCS and summarise their properties.

For real constants $\phi_0, \ n_0\geq0$ and $\sigma_0>0$, they are defined as the gaussian combination
\begin{equation}
\Psi_{\text{G}}(x,t;n_0,\sigma_0,\phi_0)=\sum_{n=1}^\infty C_n^{\text{G}}(n_0,\sigma_0,\phi_0)e^{-iE_nt}\psi_n(x),\quad C_n^{\text{G}}(n_0,\sigma_0,\phi_0)=\frac{e^{-\frac{(n-n_0)^2}{4\sigma_0^2}-in\phi_0}}{\sqrt{N_\text{G}(n_0,\sigma_0)}},\label{G}
\end{equation}
where the normalisation factor is
\begin{equation}
N_\text{G}(n_0, \sigma_0)=\sum_{n=0}^\infty e^{-\frac{(n-n_0)^2}{2\sigma_0^2}}.
\end{equation}
The resolution of the identity is satisfied for theses states, as well as time stability and continuity in $n_0$ and $\sigma_0$ \cite{Fox-Choi2000}. 

Interestingly, not only \eqref{G} has a gaussian distribution in $n$, the sum can be carried out approximately to yield a gaussian \textit{wavefunction}. Indeed, as demonstrated in \cite{FH},
\begin{equation}
\Psi_{\text{G}}(n_0,\sigma_0,\phi_0;x,t)\simeq\frac{1}{(\sqrt{2\pi}s)^{1/2}}\exp{\left[-\frac{(x-X)^2}{4s^2}+iPx\right]},\label{closedformG}
\end{equation}
(up to a $x$-independent phase factor), for $x\in[0,\pi]$, $t>0$ and with the following shorthands:
\begin{align*}
&X=\phi_0+\frac{Pt}{1/2}, \quad P= n_0,\quad
s=\frac{1}{2\sigma}, \quad \sigma^2=\frac{\tau}{4(\tau^2+t^2)}, \quad \tau=(4\sigma_0^2)^{-1}.
\end{align*}
Here ``$\simeq$'' stresses an approximate validity under the conditions $n_0\gg\sigma_0\gg1$, $X\gg s$, $\pi-X\gg s$, and $t\ll\tau$.

This result implies, as alluded to earlier, that the state is a well localised wave packet bouncing back and forth on the walls of the well, in a quasi-classical fashion. Its initial width is set by the parameter $\sigma_0$, which also determines the time scale $\tau$ of the Lorentzian decay of the packet. On the other hand, the velocity of the motion scales as the parameter $n_0$. Finally, the phase $\phi_0$ establishes the initial position of the wave packet. Here, of course, we naturally identify $X\simeq\left\langle x \right\rangle$ and $P\simeq\left\langle p \right\rangle$.

Another remarkable consequence of \eqref{closedformG} is that it explains why the GCS almost saturate Heisenberg uncertainty relation, when the conditions of validity are satisfied \cite{FH}. This is readily understood by recalling the standard fact that the most general wave function minimizing $\Delta x\Delta p$ is \cite{CT}
\begin{equation}
\Psi(x,t)=A(t)e^{-\frac{(x-\left\langle x \right\rangle)^2}{4s^2}+i\left\langle p \right\rangle x}.
\end{equation}

\subsection{Gaussian Klauder coherent states for the SUSY partners}

The construction of the GCS can easily be attempted for the SUSY partner in just the same way it was performed in the original system:
\begin{equation}
\tilde{\Psi}_{\text{G}}(x,t;n_0,\sigma_0,\phi_0)=\sum_{n=1}^\infty C_n^{\text{G}}(n_0,\sigma_0,\phi_0)e^{-iE_nt}\tilde{\psi}_n(x).\label{Gtwidle}
\end{equation}

However, the algebraic construction reviewed in section 2.1 a priori does not enforce \eqref{Gtwidle} to retain the well-understood quasi-classical behaviour summarised in \eqref{closedformG}.

We notice on the other hand that the SUSY stationary states \eqref{eigenSUSYIW} with $n\neq k$, can be regarded as perturbed versions of the initial eigenmodes \eqref{eigenIW}, provided the overall sign is chosen as in \eqref{psi_n}. This is especially obvious at $k\gg 1$, i.e. by using a high energy mode for the SUSY mapping.

The case $n=k$ deserves here, like in section 2.1, some special care. As clear from \eqref{psi_k}, the perturbative equivalence $\psi_n(x)\simeq\tilde{\psi}_n(x;k\gg1,\omega)$ no longer holds for this state. Thus, the mode $\tilde{\psi}_k(x;k,\omega)$ corresponding to the energy used for the SUSY mapping is always fundamentally different from its original homologue $\psi_k(x)$.

The consequences of these remarks for the GCS now appear more distinctly. Whenever $n_0$ is chosen far (i.e. out of a few standard deviations) from $k$, the GCS of the SUSY partner agree with the GCS of the original system. The analysis of section 2.2 then carries on, and a localised gaussian quasi-classical wave packet minimizing approximately Heisenberg relation is generated.

If $n_0$ and $k$ are of the same order, on the other hand, some discrepant features are expected for the SUSY GCS. As a concrete example, figure 2 exhibits the time-evolution of the probability density distribution calculated from \eqref{Gtwidle}, in the \textit{coincident} case $n_0=k$. Clearly, an effect of the supersymmetry is to generate superimposed wavelets on the principal gaussian wave packet. As time evolves, moreover, some side modes appear next to the dominant packet. Numerical calculations show that these ``harmonics'' coherently follow the principal peak in its displacements inside the well as time evolves. The wave packet slowly spreads and bounces on the walls as in the original construction.

\begin{center}
\includegraphics[scale=0.51]{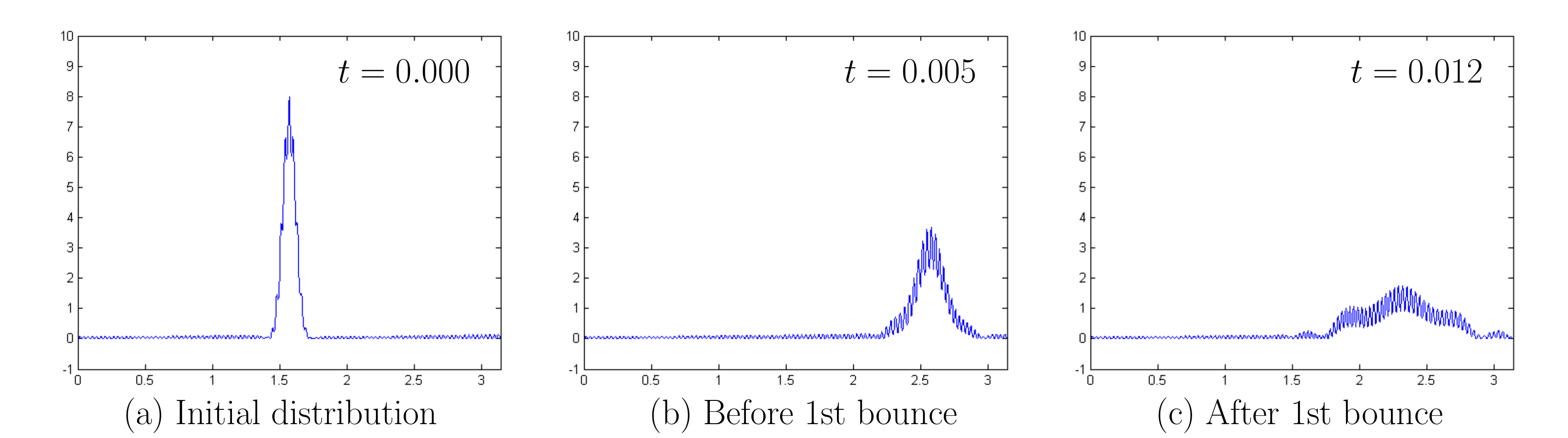}

\small{Figure 2 - Time-evolution of the probability density distribution inside the well for the SUSY coherent state $\tilde{\Psi}_{\text{G}}(x,t;n_0=100,\sigma_0=10,\phi_0=\pi/2)$. The SUSY parameters are: $k=100$, $\omega=2$.}
\end{center}

Let us now turn to the interesting question of the effect of the SUSY parameter $\omega$ on the GCS. Our numerical calculations seem to suggest that the magnitude of $|\omega-1/2|$ weakly affect the behaviour of the states. Thus, the above discussion continues to apply. What is more subtle is the effect of changing the sign of $(\omega-1/2)$. In spite of its simple mirror effect on the potential, as discussed in section 2.1, it tremendously affects the GCS. As shown on figure 3, the $\mathbb{Z}_2$ action suppresses the wavelets and side-modes, even for the coincident case $n_0=k$. This is surprising since the above discussion is still applicable, so we might still have expected some disagreement due to the exceptional state $\tilde{\psi}_k(x;k,\omega)$. At the level of the wave functions \eqref{psi_n} and \eqref{psi_k}, the $\mathbb{Z}_2$ action acts as an alternating phase factor (except at energy $E_k$, where the alternating pattern is broken): $\tilde{\psi}_n\rightarrow-\cos(n\pi)\tilde{\psi}_n$, $\tilde{\psi}_k\rightarrow+\cos(k\pi)\tilde{\psi}_k$ (provided we also map $x\rightarrow\pi-x$).

Note that although the GCS of the SUSY Hamiltonian can be chosen to be almost identical to the GCS of the original system, the potential functions remain very different. Indeed, the limit $k\gg 1$ emphatically does not lead to $\tilde{V}(x;k,\omega)\simeq 0$ in $[0,\pi]$. Thus, the construction of well behaved coherent states for this nontrivial system is a remarkable consequence of the SUSY approach.

\begin{center}
\includegraphics[scale=0.51]{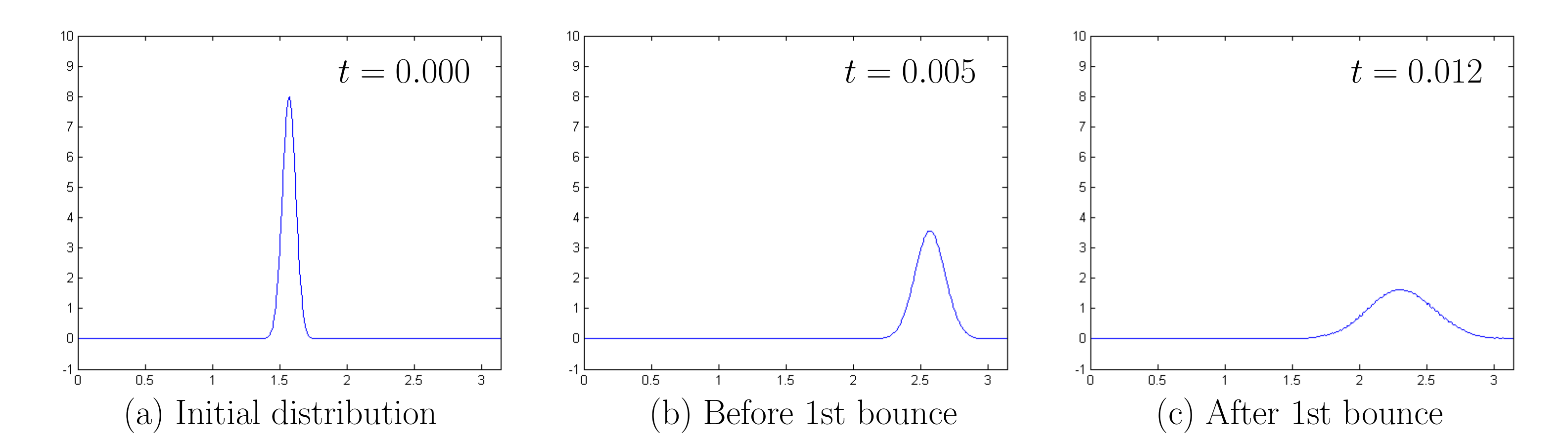}

\small{Figure 3 - Time-evolution of the probability density distribution inside the well for the SUSY coherent state $\tilde{\Psi}_{\text{G}}(x,t;n_0=100,\sigma_0=10,\phi_0=\pi/2)$. The SUSY parameters are: $k=100$, $\omega=-1$.}
\end{center}

As a concluding side comment, let us note that the classical dynamics in the SUSY system \eqref{potential}, which is generally expected for coherent state constructions, is not immediately transparent. A minimal criteria for classicality might thus be chosen to be a relatively slowly evolving and highly localized wave envelope. We have adhered to this point of view in the current work.

\section{The 2D infinite well and SUSY partners }

\subsection{Description of the original model}

We consider again a particle of mass 1/2, now assumed to move in a square 2D box of size $\pi$. The Hamiltonian is given as
\begin{equation}
H= H_x+H_y,
\end{equation}
where $H_x$ and $H_y$ are the 1D Hamiltonians of the infinite well in the directions $x$ and $y$ respectively. The corresponding normalised eigenstates and discrete energies are
\begin{equation}
 \Psi_{n,m}(x,y)=\psi_n(x)\psi_m(y)=\frac{2}{\pi}\sin nx \sin my
\end{equation}
and 
\begin{equation}
E_{n,m}= n^2+m^2,
\label{en2D}
\end{equation}
where $n,m=1,2,\dots$

We see that this quadratic energy spectrum presents two types of degeneracies. The first type is a permutation degeneracy since the eigenstates $\Psi_{n,m}(x,y)$ and $\Psi_{m,n}(x,y)$, $n\neq m$, are distinct with the same energy $E_{n,m}$. The second type is called accidental  \cite{Fox-Choi2001, DH} or arithmetic \cite{Itzykson}. For example, we have $E_{5,5}=E_{1,7}=50$ with $\Psi_{5,5}(x,y)$ and $\Psi_{1,7}(x,y)$ as distinct eigenfunctions and  $E_{1,8}=E_{4,7}=65$ with $\Psi_{1,8}(x,y)$ and $\Psi_{4,7}(x,y)$ as distinct eigenfunctions. The problem of identifying the number and type of these degeneracies for the infinite well has already been considered  \cite{Fox-Choi2001, DH, Itzykson}.

We present here some notations that will be useful in the discussion of coherent states of such a system:
\begin{itemize}
\item We rank the energies \eqref{en2D} in increasing order and write them as $E_\nu$, with index $\nu=0,1,\dots$ We also introduce the shifted energies ${{\mathcal E}_\nu}=E_\nu-E_0$.

\item As alluded to above, more than one state may correspond to $E_\nu$. We thus define $\mu_\nu$ as the index for the summation of all the eigenstates associated to the degenerate energy $E_\nu$ and $d_\nu$ as the associated number of degeneracies. The index $\mu_\nu$ goes from 0 to $d_\nu-1$. Although they are irrelevant for our purposes, let us point out that explicit formula for the existence and number of accidental degeneracies exist in the number theoretic literature. They involve the number of divisors of $E_\nu$ of the form $4k+1$ and $4k+3$, $k\in\mathbb{N}$.

\item The indices $(n,m)$ corresponding to a state alternatively referred to with $(\nu,\mu_\nu)$ are written $(n_{\nu,\mu_\nu},m_{\nu,\mu_\nu})$.

\item We rename the eigenstate $\Psi_{n_{\nu,\mu_\nu},m_{\nu,\mu_\nu}}(x,y)$, the state corresponding to the energy $E_\nu$ and degeneracy index $\mu_\nu$, as $\Psi^{\nu,\mu_\nu}(x,y)$.

\item To each energy $E_\nu$, we can thus associate a single ``cumulative'' eigenstate constructed as a superposition of the states  $\Psi^{\nu,\mu_\nu}(x,y)$. It takes the form
\begin{equation}
\Phi^{\nu}(x,y)=\sum_{\mu_\nu=0}^{d_{\nu}-1}\gamma_{\nu,\mu_\nu} \Psi^{\nu,\mu_\nu}(x,y),
\label{superpositionphi}
\end{equation}
where the coefficients $\gamma_{\nu,\mu_\nu}$ are arbitrary non zero complex numbers. Note that we have taken into account the permutation degeneracy in the superposition formula.

\item Finally, a polar parametrization ($\rho_\nu=\sqrt{E_\nu}$, $\tan{\theta_{\nu,\mu_\nu}}=\frac{n_{\nu,\mu_\nu}}{m_{\nu,\mu_\nu}}$) is introduced. It provides an intuitive picture of the indices $\nu$ and $\mu_\nu$. As exemplified in figure 4, the energy index $\nu$ is associated with the \textit{radius} $\rho_\nu$ in $(m,n)$-plane, while the degeneracy index $\mu_\nu$ is associated with the \textit{angle} $\theta_{\nu,\mu_\nu}$. For definiteness, let us assume that $\mu_\nu$ increases as $n_\nu$ increases, while $m_\nu$ decreases (for fixed $\nu$). Put differently, $\mu_\nu$ increases as the angle $\theta_{\nu,\mu_\nu}$ increases.

\end{itemize}

\begin{figure}
\begin{center}
\includegraphics[width=.51\textwidth]{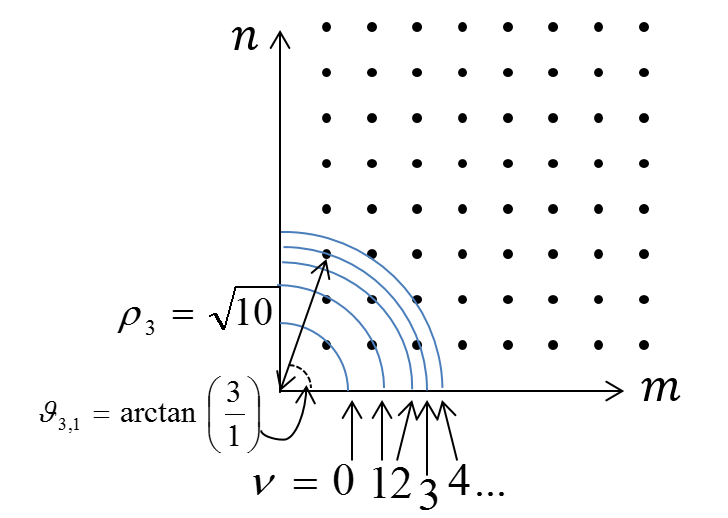}

\small{Figure 4 - Polar parametrisation of eigenstates}
\end{center}
\end{figure}

\subsection{Construction of SUSY partners}

2D generalisations of SUSY quantum mechanics have been proposed some time ago and, in particular, the approach of Ioffe and collaborators \cite{Ioffe} has given some relevant results. Unfortunately, using supercharges that are essentially a linear combination of the ones used in 1D does not lead to the new potentials we have already obtained in 1D. An alternative approach is considered here. 

A SUSY partner Hamiltonian which is separable in $x$ and $y$ could be obtained from SUSY 1D Hamiltonians as constructed in Section 2. Indeed, let us take $Q=Q_x Q_y$ and $Q^\dagger= Q_{x}^\dagger Q_{y}^\dagger$ where $Q_x, \ Q_{x}^\dagger$ and  $Q_y, \ Q_{y}^\dagger$ satisfy the interwining 1D relations (\ref{intertwining}) in the corresponding variables.  We thus easily get
\begin{equation}
\tilde{H} Q^\dagger=Q^\dagger H,\quad Q \tilde{H}  = H Q,
\label{intertwining2D}
\end{equation}
where $\tilde{H} =\tilde{H}_{x}+ \tilde{H}_{y}$ with $\tilde{H}_{x}=- \frac{d^2}{dx^2}+V(x,k_1,\omega_1)$ and $\tilde{H}_{y}=- \frac{d^2}{dy^2}+V(y,k_2,\omega_2)$. The corresponding potentials are explicitly given in (\ref{potential}).

Eigenstates of the SUSY partner Hamiltonian $\tilde{H}$ are thus easily constructed from
\begin{equation}
\tilde{\Psi}_{n,m}(x,y;k_1,k_2,\omega_1,\omega_2)= ((k_1^2-n^2)(k_2^2-m^2))^{-1}Q^+\Psi_{n,m}(x,y)=\tilde{\psi}_n(x;k_1,\omega_1)\tilde{\psi}_m(y;k_2,\omega_2),
\label{eigenSUSYIW2D}
\end{equation}
for $n\neq k_1,m\neq k_2$. The corresponding energies are $E_{n,m}=(\ref{en2D})$.
In 1D, we had to add one state (\ref{psi_k}) in order to get a complete spectrum. In 2D, we have to add more states. Indeed, we also get the following states:
\begin{equation}
\tilde{\Psi}_{k_1,m}(x,y;k_1,k_2,\omega_1,\omega_2)= \tilde{\psi}_{k_1}(x;k_1,\omega_1)\tilde{\psi}_m(y;k_2,\omega_2),
\label{eigenSUSYIW2Dk1}
\end{equation}
with energy $E_{k_1,m}$,
\begin{equation}
\tilde{\Psi}_{n,k_2}(x,y;k_1,k_2,\omega_1,\omega_2)=\tilde{\psi}_n(x;k_1,\omega_1)\tilde{\psi}_{k_2}(y;k_2,\omega_2),
\label{eigenSUSYIW2Dk2}
\end{equation}
with energy $E_{n,k_2}$,
and finally the eigenstate with energy $E_{k_1,k_2}$ is given as
\begin{equation}
\tilde{\Psi}_{k_1,k_2}(x,y;k_1,k_2, \omega_1,\omega_2)= \tilde{\psi}_{k_1}(x;k_1,\omega_1)\tilde{\psi}_{k_2}(y;k_2,\omega_2).
\label{eigenSUSYIW2Dk1k2}
\end{equation}

\subsection{Coherent states}
Generalised and gaussian coherent states have been constructed for the usual 2D infinite well \cite{Fox-Choi2001, DH}. The generalised coherent states (GeCS) are as usual defined as eigenstates of an annihilation operator of the system under consideration. But in the 2D case, due to the existence of a quadratic degenerate energy spectrum, we had to express the energies in increasing order and also to make a superposition of the different eigenstates with the same energy as given in (\ref{superpositionphi}).  
The construction of an annihilation operator and numerical calculations giving the behaviour of those states can be found in \cite{DH}. They have been shown to be closely related to GCS for some values of the coherent states parameters.

We will slightly adjust these constructions in this work. 

\subsubsection {Gaussian coherent states}

The definition of GCS is proposed as a direct generalisation of the 1D case. Indeed, we take
\begin{equation}
\Psi_{\text{G}}^{2D}(x,y,t; n_0,m_0,\sigma_{n_0},\sigma_{m_0},\phi_{n_0},\phi_{m_0})=\Psi_{\text{G}}(x,t;n_0,\sigma_{n_0},\phi_{n_0})\Psi_{\text{G}}(y,t;m_0,\sigma_{m_0},\phi_{m_0}),
\end{equation}
with $\Psi_{\text{G}}=(\ref{G})$. Explicitly, we get 
\begin{equation}
\Psi_{\text{G}}^{2D}(x,y,t; n_0,m_0,\sigma_{n_0},\sigma_{m_0},\phi_{n_0},\phi_{m_0})=\sum_{n=1}^\infty \sum_{m=1}^\infty C_{n,m}^{\text{G}}e^{-i E_{n,m}t} \Psi_{m,n}(x,y),
\label{gcs2d}
\end{equation}
where
\begin{equation}
C_{n,m}^{\text{G}}=C_{n,m}^{\text{G}}(n_0,m_0,\sigma_{n_0},\sigma_{m_0},\phi_{n_0},\phi_{m_0})=C_{n}^{\text{G}}(n_0,\sigma_{n_0},\phi_{n_0}) C_{m}^{\text{G}}(m_0,\sigma_{m_0},\phi_{m_0}).
\end{equation}

This definition pays off since the analysis of \cite{FH} reviewed in section 2.2 continues to apply here. As in the 1D case, we get a well localised state and quasi-classical behaviour. Approximate saturation of the Heisenberg uncertainty bound is also achieved, as in the 1D setting.

\begin{center}
\includegraphics[width=.7\textwidth]{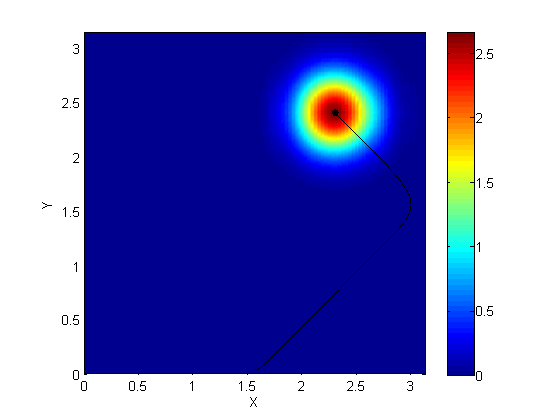}

\small{Figure 5 - Probability density distribution and time evolution of the position expectation value of the SUSY GCS in 2D until $t=0.012$. State parameters are chosen as: $n_0=m_0=100$, $\sigma_{n_0}=\sigma_{m_0}=10$, $\phi_{n_0}=\pi/2$, and $\phi_{n_0}=0$. SUSY parameters are: $k_1=100$, $k_2=50$,  $\omega_1=-1$, and $\omega_2=2$.}
\end{center}

Such a definition of GCS is easily extended to the SUSY case where the eigenstates $\Psi_{m,n}(x,y)$ in (\ref{gcs2d}) are replaced by the eigenstates $\tilde\Psi_{m,n}(x,y)$. Again, the factorization of the state yields a straightforward generalisation of the 1D results. As an example, figure 5 shows a trace of the time evolution of the position expectation value for the SUSY GCS in 2D. A neat localized packet bouncing on the walls is obtained once again.

\subsubsection{Generalised coherent states}

The GeCs are eigenstates of an annihilation operator of the 2D infinite well and they have been defined as \cite{DH}  
\begin{equation}
\Psi_{\text{Ge}} (x,y,t ;z)=\frac{1}{\sqrt{N_{\text{Ge}}(z)}} \sum_{\nu=0}^{\infty} \frac{z^\nu}{\sqrt{\rho(\nu)}}e^{-i {{\mathcal E}_\nu}t }\Phi^{\nu}(x,y),
\label{gecs}
\end{equation}
with the normalisation factor
\begin{equation}
N_\text{Ge}(z)\equiv\sum_{\nu=0}^\infty \frac{|z|^{2\nu}}{\rho(\nu)}\nonumber 
\end{equation}
and 
\begin{equation}
\rho(\nu)=\begin{cases} \ 1,&\text{if}\ \nu=0 \\\ {\Pi}_{i=1}^{\nu} {\mathcal E}_i ,&\text{if}\  \nu\neq 0.\end{cases}
\end{equation}
These states depend on a continuous complex parameter $z$ as in the 1D case. With respect to an alternative approach \cite{DH}, more freedom is given to the states $\Phi^{\nu}(x,y)$ being a superposition of states with same energy ${{\mathcal E}_\nu}$.

\section{Conclusion and future work}

SUSY partners of the infinite well have been constructed and have shown to satisfy relevant properties compared with the ones of the original quantum system. In particular, for the 1D system, we have constructed a set of coherent states (GCS) which depends on real and discrete parameters. The relation between those states and the generalised ones (GeCS) has been formally given \cite{FH} and we have shown the behaviour of the GCS case.

In the 2D case, we have extended the construction of SUSY partners to get similar potentials as in 1D. More parameters are involved in this context and the existence of degeneracies in the energy spectrum has lead us to adjust the definition of coherent sates  with respect with the ones used in the 1D case. The GCS are constructed in order to have a good behaviour with respect to localisation in the usual case as well as in the SUSY case. We have also constructed GeCS in the 2D setting. 

It remains to make a link between both types of states. The use of the polar parametrisation is clearly a way to solve the problem. Indeed, the GCS may be written as

\begin{equation}
\Psi_{\text{G}}(x,y,t)\equiv\frac{1}{\sqrt{N_\text{G}}}\sum_{\nu=0}^\infty {e^{\frac{(\rho_\nu-{\bar \rho}_0)^2}{4\sigma_0^2}} e^{-i {\mathcal{E}_\nu} t}} \Psi^{\text{G},\nu}(x,y).
\end{equation}
The gaussian terms depending on the radial variables could thus be related to the factor $z^\nu({\rho(\nu)})^{-\frac12}$ in (\ref{gecs})
as it was the case in 1D \cite{FH}. The gaussian terms involving the angular variables are related to the coefficients of the superposition of eigenstates with same energy. This way, the states $\Psi^{G,\nu}(x,y)$ will be related to $\Phi^{\nu}(x,y)$ in (\ref{gecs}). As mentioned earlier, a particular approach of this question in 2D \cite{Fox-Choi2001,DH} has shown that GeCS are in fact a good approximation of  GCS. In a future work, we hope to formally solve this problem.

\section*{Acknowledgements}
This work has been supported in part by research grants from Natural sciences and engineering research council of Canada (NSERC). Marc-Antoine Fiset acknowledges a NSERC fellowship.

\section*{References}
%%%%%%%%%%%%%%%%%%%%%%%%%%%%%%%%%%%%%%%%%%%

\end{document}